\documentclass{appolb}
\usepackage{epsfig}

\begin{document}
\title{Handling Excited States on the Lattice: \\
The GEVP Method~\thanks{Presented at Excited QCD 2010.}%
}
\author{Tereza Mendes
\address{Instituto de F\'isica de S\~ao Carlos, Universidade de S\~ao Paulo,\\
Caixa Postal 369, 13560-970 S\~ao Carlos, SP, Brazil
}
}
\maketitle
\begin{abstract}
High-precision calculations of hadron spectroscopy are a
crucial task for Lattice QCD. State-of-the-art techniques are needed
to disentangle the contributions from different energy states, such as
solving the generalized eigenvalue problem (GEVP) for zero-momentum
hadron correlators in an efficient way. We review the method and
discuss its application in the determination of the $B_s$-meson spectrum
using (quenched) nonperturbative HQET at order $1/m_b$.
\end{abstract}
  
\section{Introduction}

Systematic errors in lattice simulations include
not only finite-size and discretization effects, dependence 
on the chiral extrapolation and quenching, but also
systematic effects due to contamination from higher excited states
in the calculation of energy levels (masses) of quark bound states.
This is because masses and decay constants are computed in lattice 
QCD from the exponential decay of Euclidean correlation functions 
$C(t)$, which are built from composite fields with the quantum numbers
of a given state (see e.g.\ \cite{Gattringer:2007da}).
More precisely, in the simulation, one evolves gluon fields (the link 
variables $U$) in the Monte Carlo dynamics associated with the partition 
function
\begin{eqnarray}
Z&=&\int {\cal D}U \,e^{- S_g}\!
\int {\cal D}\psi \, {\cal D}{\overline \psi}
\;\, e^{\,-\!\int d^4x\; {\overline \psi}(x)\, K\,\psi(x) }
\nonumber \\[2mm]
&=& \int {\cal D}U \,e^{-S_g}\,\det K(U)\,,
\end{eqnarray}
where $S_g$ is the gluonic action and $K(U)$ is the Dirac operator.
(The quenched approximation corresponds to $\det K = 1$.) Although the
quark fields $\psi$ are not evolved directly,
information about them may be obtained once the link configuration is 
produced, since quark propagators are given by 
$\langle\psi \, {\overline \psi}\rangle \,=\, 
\langle K^{-1}\rangle \,,$ i.e.\ by computing
the inverse of $K(U)$ for each configuration of the link variables
and averaging over the configurations produced.
Similarly, the computed inverse of $K(U)$ may be used to build interpolators, 
i.e.\ products of creation and annihilation operators with the quantum 
numbers of the desired bound state and a good overlap with the hadron 
wave function on the lattice. The result is then averaged over the
configurations to yield the (Euclidean) correlators $C(t)$ as
\begin{equation}
C(t)\;=\;\sum_{{\bf x}} \langle J({\bf x},t)\,J(0,0)\rangle 
\;\equiv\;\langle O(t)\,O(0)\rangle , 
\label{eq:corr}
\end{equation}
where $\,J({\bf x},t)={\overline \psi}\,\Gamma \,\psi\,$ and $\Gamma$ is the
appropriate Dirac matrix (e.g.\ $\,\Gamma=\gamma_5\,$ for
pseudoscalar mesons).
Finally, one determines masses and decay constants by identifying the
various contributions to the spectral decomposition
\begin{equation}
\langle O(t)\,O(0)\rangle  \;=\; C(t) \;=\; 
\sum_{n=1}^{\infty}\, |\langle n|\,\hat{O}\,|0 \rangle |^2\; e^{-E_n t}\,,
\label{eq:corr2}
\end{equation}
where $|n\rangle$ are eigenstates of the Hamiltonian (i.e.\ the logarithm of 
the transfer matrix) and all energies $E_n$ have the vacuum energy subtracted.
Also, we assume Hermitean operators and a large enough time extent of the 
lattice to yield the simple exponential form above. 
At large $t\;$ we expect to observe a plateau in the ``effective mass''
\begin{equation}
E_1^{\rm eff}(t)\;\equiv\;\log [C(t)/C(t+1)]
\,\;\to\;\, E_1 \,+\,{\cal O}({e}^{-(E_2-E_1)t})\,,
\end{equation}
in such a way that the true ground-state energy $E_1$ may be estimated.

Clearly, determining masses and decay constants from the correlators in 
Eq.\ (\ref{eq:corr2}) is not an easy task.
To see this, consider the first correction above, given by
$\,\exp\left[-(E_2-E_1)\,t\right]\,$ with a positive coefficient, 
in the case of a heavy-light system. 
For typical energy differences of a few hundred MeV, 
a plateau can only be reached for $t$ around 1 fm, but by then the signal 
has started to compete with noise, even for improved static-quark 
discretizations. 
This is a general problem 
and trying to determine subleading corrections by multiple-exponential 
fits leads to large systematic errors. One alternative is to use more 
sophisticated fitting methods, such as Bayesian fitting, evolutionary 
fitting and NMR-inspired methods. Another way to ensure better precision
is inspired by the variational method in quantum mechanics and
consists in increasing the basis of interpolators to build a matrix
of correlators $C_{ij}(t)$, for which a Generalized Eigenvalue Problem 
(GEVP) is formulated (see e.g.\ \cite{Luscher:1990ck}).
One then considers all-to-all propagators \cite{Foley:2005ac}
instead of simple point sources as indicated in Eq.\ \ref{eq:corr}.

The GEVP is a valuable tool to reduce systematic errors in 
the above determinations, and thus to deliver high-precision tests of QCD.
We summarize the derivation of an optimal use of the method in Section
\ref{method} below and describe its application to spectrum
calculations of $B_s$ mesons in nonperturbative HQET to order $1/m_b$
in Section \ref{HQET}.

\section{The Method}
\label{method}

The GEVP is defined by
\begin{equation}
C(t)\,v_n(t,t_0)\;=\;
\lambda_n(t,t_0)\,C(t_0)\,v_n(t,t_0)\,,
\end{equation}
where $t>t_0$ and $C(t)$ is now a matrix of correlators, given by
\begin{equation}
C_{ij}(t)\,\;=\;\,\langle O_i(t) O_j(0)\rangle
\,\;=\;\, \sum_{n=1}^\infty {e}^{-E_n t}\, \Psi_{ni}\Psi_{nj}\,,\quad
i, j = 1,\ldots, N\,.
\end{equation}
The chosen interpolators $O_i$ are taken (hopefully)
linearly independent, e.g.\ they may be built from
smeared quark fields using $N$ different smearing levels.
The matrix elements $\Psi_{ni}$ are defined by
\begin{equation}
\Psi_{ni} \;\equiv\; (\Psi_n)_i \;=\; \langle n|\hat O_i|0\rangle\;,
\quad\; \langle m|n\rangle \,=\,\delta_{mn}\,.
\end{equation}

One thus computes $C_{ij}$ for the interpolator basis $O_i$ from
the numerical simulation, then gets effective energy levels $E_n^{\rm eff}$
and estimates for the matrix elements $\Psi_{ni}$ from the
solution $\lambda_n(t,t_0)$ of the GEVP at large $t$. For the energies
\begin{equation}
E_n^{\rm eff}(t,t_0)\;\equiv\;
{1\over a} \, \log{\lambda_n(t,t_0) \over \lambda_n(t+a,t_0)}
\end{equation}
it is shown \cite{Luscher:1990ck}
that $E_n^{\rm eff}(t,t_0)$ converges exponentially as $t\to\infty$
(and fixed $t_0$) to the true energy $E_n$. However, since the
exponential falloff of higher contributions may be slow,
it is also essential to study the convergence as a function of
$t_0$ in order to achieve the required efficiency for the method.
This has been done in \cite{Blossier:2009kd}, by explicit application
of (ordinary) perturbation theory to a hypothetical truncated problem
where only $N$ levels contribute. The solution in this case is exactly
given by the true energies, and corrections due to the higher states
are treated perturbatively. We get
\newcommand{\aeff}{\hat {\cal A}_n^{\rm eff}}
\newcommand{\qeff}{\hat {\cal Q}_n^{\rm eff}}
\newcommand{\corren}{\varepsilon_{n}}
\newcommand{\corrpn}{\pi_{n}}
\begin{equation}
E_n^{\rm eff}(t,t_0) \;=\;
E_n \,+\, {\varepsilon_{n}(t,t_0)}\,
\end{equation}
for the energies and
\begin{equation}
{e}^{-\hat H t}(\qeff(t,t_0))^\dagger|0\rangle \;=\;
|n\rangle \,+\, \sum_{n'=1}^\infty  \pi_{nn'}(t,t_0)
\, |n'\rangle
\end{equation}
for the eigenstates of the Hamiltonian, which may be estimated
through
\begin{eqnarray}
    \qeff(t,t_0) &=& R_n \,(\hat O\,,\,v_n(t,t_0)\,) \,, \\[2mm]
R_n &=& {\left(v_n(t,t_0)\,,\, C(t)\,v_n(t,t_0)\right)}^{-1/2}
\; \left[{\lambda_n(t_0+a,t_0) \over \lambda_n(t_0+2a,t_0)}\right]^{t/2}\,.
\end{eqnarray}

\begin{figure}[t]
\begin{center}
\vskip 0mm
\includegraphics[width=0.3\textwidth]{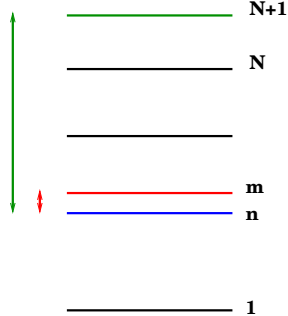}
\vskip 5mm
\caption{Schematic representation of energy levels,
showing how a solution of the GEVP for conveniently
chosen $t$, $t_0$ yields conversion to the asymptotic state
controlled by a much larger energy gap than usual (i.e.\ the one
represented by the longer arrow instead of the shorter one).}
\label{levelsplot}
\vspace{-8mm}
\end{center}
\end{figure}

In our analysis we see that, due to cancellations of $t$-independent
terms in the effective energy, the first-order corrections in
${\varepsilon_{n}(t,t_0)}$ are independent of $t_0$ and very strongly
suppressed at large $t$. We identify two regimes:
1) for $t_0 \,<\, t/2$, the 2nd-order corrections dominate and
their exponential suppression is given by the smallest energy gap
$\,|E_m-E_n|\equiv \Delta E_{m,n}\,$ between level $n$ and its neighboring
levels $m$; and 2) for $t_0 \,\geq\, t/2$,
the 1st-order corrections dominate and the suppression is given by the
large gap $\Delta E_{N+1,n}$.
Amplitudes $\,\pi_{nn'}(t,t_0)\,$ get main contributions
from the first-order corrections. For fixed $t-t_0$ these are also
suppressed with $\Delta E_{N+1,n}$.
Clearly, the appearance of large energy gaps in the second regime
improves convergence significantly. A pictorial illustration of the
improvement in shown in Fig.\ \ref{levelsplot}.
We therefore work with $t$, $t_0$ combinations in this regime.


\def\first{1/m_b}
\def\stat{\rm {stat}}

A very important step of our approach is to realize that the same
perturbative analysis may be applied to get the leading
corrections to correlators in an effective theory, such as
corrections of order $1/m_b$ to the static case in HQET correlation 
functions. These are given by
\begin{equation}
  C_{ij}(t) \;=\; C_{ij}^{\stat}(t) \,+\,
  \omega \,C_{ij}^{\first}(t) \,+\, {\cal O}(\omega^2)\,,
\end{equation}
where the combined ${\cal O}(1/m_b)$ corrections are symbolized by
the expansion parameter $\omega$.
Following the same procedure as above, we get similar exponential
suppressions (with the static energy gaps) for static and ${\cal O}(1/m_b)$
terms in the effective theory. We arrive at
\begin{equation}
    E_n^{\rm eff}(t,t_0) \;=\; E_n^{{\rm eff},{\rm stat}}(t,t_0)
     +\omega E_n^{{\rm eff},{1/m_b}}(t,t_0) +{\cal O}(\omega^2)
\end{equation}
with
\begin{eqnarray}
    E_n^{{\rm eff},{\rm stat}}(t,t_0) &=&
    E_n^{\stat} \,+\, \beta_n^{\stat} \,
    {e}^{-\Delta E_{N+1,n}^{\stat}\, t}+\ldots\,,
\label{effenergies}
\\[2mm]
    E_n^{\rm eff,\first}(t,t_0) &=&
    E_n^{\first} \,+\, [\,\beta_n^{\first}
    \,-\, \beta_n^{\stat}\,t\;\Delta E_{N+1,n}^{\first}\,]
    {e}^{-\Delta E_{N+1,n}^{\stat}\, t}+\ldots\, .
\end{eqnarray}
and similarly for matrix elements.

An application of the methods described in this section is discussed
next. 

\section{Application to Nonperturbative HQET}
\label{HQET}

High-precision hadronic matrix elements are a key ingredient as 
theoretical inputs in B physics and are ideally obtained from 
lattice-QCD simulations. However, currently used lattices
are not large enough to represent simultaneously the low-energy 
scale of $\Lambda_{\rm QCD}$, which requires a large 
physical lattice size, and the high-energy scale of the 
heavy-quark mass $m_b$, which requires a very small lattice 
spacing $a$. A promising alternative is to consider (lattice) 
heavy-quark effective theory (HQET), which allows for an elegant 
theoretical treatment, with the possibility of fully nonperturbative 
renormalization \cite{NPHQET}. 

HQET provides a valid low-momentum description for systems with one heavy
quark, with manifest heavy-quark symmetry in the static limit.
The heavy-quark flavor and spin symmetries are broken at finite values
of $m_b$ respectively by kinetic and spin terms, with first-order
[i.e.\ ${\cal O}(1/m_b)$] corrections to the static Lagrangian incorporated 
by an expansion of the statistical weight in $1/m_b$, such that the
symmetry-breaking operators are treated as insertions into static
correlation functions.
This guarantees the existence of a continuum limit,
with results that are independent of the regularization, provided that
the renormalization be done nonperturbatively.
Masses and decay constants are expanded as sums of a static and an
${\cal O}(1/m_b)$ contribution, in terms of the parameters of the effective
theory and of the bare energies and matrix elements, which are computed
in the numerical simulation.
The divergences (with inverse powers of $a$) in these parameters are
cancelled through the nonperturbative renormalization, which is based on
matching the HQET parameters to QCD on lattices of small physical volume
--- where fine lattice spacings can be considered --- and extrapolating to
a large volume by the step-scaling method. This analysis has been
recently completed for the quenched case \cite{Blossier:2010jk}.
Using the computed HQET parameters, we have carried out a study 
\cite{Blossier:2010vz} (see also \cite{Blossier:2009mg}) of spectrum 
and decay constants for $B_s$ mesons applying the GEVP method as
described in the previous section.

We have employed two lattice actions for the static quark,
lattices of spatial extent $L\approx1.5$ fm with three lattice spacings
and all-to-all strange-quark propagators constructed from
approximate low modes, with 100 configurations.
The interpolating fields were obtained from a simple $\gamma_0\gamma_5$ 
structure and 8 levels of Gaussian smearing for the strange-quark field.
The resulting ($8\times8$) correlation matrix may be
conveniently truncated to an $N\times N$ one and the GEVP
solved for each $N$, so that results can be studied as a function of $N$.
We have used two methods for picking a basis from the above interpolators
and checked that both yielded the same results.


The combined use of nonperturbatively determined HQET parameters
and efficient GEVP solution allowed us to reach a precision of a 
few percent in matrix elements and of a few MeV in energy levels, 
even with only a moderate number of configurations.
A corresponding study for the $N_f=2$ case is in progress.


\end{document}